\documentclass[aps,twocolumn,superscriptaddress,floatfix,longbibliography]{revtex4-1}
\usepackage{tikz}
\usepackage{lipsum}
\usepackage{graphicx}
\usepackage[export]{adjustbox}
\usepackage{amsmath,amssymb,wasysym}
\usepackage{braket}
\usepackage{mathtools}
\usepackage{mathrsfs}
\usepackage{verbatim}
\usepackage{makecell}
\usepackage{cases}
\usepackage{tikz}
\usepackage{booktabs}
\usepackage{slashed}
\usepackage{hyperref}
\usepackage{xcite}
\usepackage{xr}

\usepackage{xcolor}
\usepackage{pgfplots}
\makeatletter
\newcommand{\im}{\mathbf{i}}
\newcommand{\Id}{\mathbb{I}}

\newcommand{\gammaa}{\pmb{\gamma}}

\newcommand{\cS}{\mathcal{S}}
\newcommand{\cP}{\mathcal{P}}
\newcommand{\cs}{\textbf{s}}
\newcommand{\cp}{\textbf{p}}

\newcommand{\cM}{\mathcal{M}}
\newcommand{\cm}{\textbf{m}}
\newcommand{\cJ}{\mathcal{J}}
\newcommand{\cj}{\textbf{j}}

\newcommand{\x}{\textbf{x}}
\newcommand{\trs}{\mathcal{T}}

\begin{document}
\title{Instabilities of a Generalized Gross-Neveu-Yukawa Quantum Criticality}
	
	\author{Jaewon Kim}
	\affiliation{Department of Physics, University of California, Berkeley, CA 94720, USA}
    \affiliation{Department of Physics and Anthony J. Leggett Institute for Condensed Matter Theory, University of Illinois Urbana-Champaign, Urbana, Illinois 61801, USA}    
	
	\begin{abstract}
    We study the instabilities to the conformal critical point of an exactly solvable family of Gross–Neveu-Yukawa models.
    Using conformal field theory techniques, we construct the zero-temperature phase diagram and identify the superconducting and charge neutral ordered phases that destabilize the critical point.
    Both instabilities appear only when the fermions are strongly renormalized, above a critical anomalous dimension.
    A higher fermion anomalous dimension also raises the critical degree of time-reversal-symmetry breaking required to suppress superconductivity, indicating that pairing becomes more robust with stronger renormalization.
	\end{abstract}
	
	\maketitle

\section{Introduction}
Twisted van der Waals heterostructures have opened a new frontier in quantum materials, where narrow moiré bands and strong electron correlations give rise to a wealth of phenomena previously unseen in solid-state experiments \cite{MATBG1,Ins_MATBG,MATBG_SM,MATBG_SM2,FCI_0,FCI_1,FCI_2,WSe2_sc1,WSe2_sc2,rhombo_chiralsc,rhombo_fqh,rel_mott}.
Among these is the recent observation of a relativistic Mott transition in twisted WSe$_2$, where interacting Dirac fermions spontaneously break chiral symmetry and acquire a mass gap \cite{rel_mott}.
Such transitions, long theorized in relativistic fermion models, had remained inaccessible in ordinary materials due to insufficient interaction strength.
In moiré WSe$_2$, however, tuning the twist angle continuously lowers the Dirac velocity, enabling access to this relativistic Mott regime.

The low-energy theory of this transition is described by the Gross–Neveu-Yukawa (GNY) model \cite{Graphene_Herbut1,Graphene_Herbut2,GNXY_Hawashin}, yet its analytical characterization at strong coupling remains notoriously challenging.
While recent numerical approaches—such as conformal bootstrap analyses~\cite{Iliesiu_2018,Erramilli_2023} and quantum Monte Carlo simulations~\cite{GN_QMC0,GN_QMC,GN_QMC2,GN_QMC3,GN_QMC4,GN_QMC5}—have provided valuable insights into Gross–Neveu criticality, analytic approaches to this critical point have largely relied on perturbative techniques, such as the $\epsilon$-expansion and the conventional $1/N$ expansion \cite{GN_org,GNZ2_phasediagram,GNXY_LargeN,GN_2p,GN_4m,Herbut_2024,Graphene_Herbut1,Graphene_Herbut2,GNY_SO2N1,GNY_SO2N2,GNXY_Hawashin,GNSO4_Uetrecht}.
In particular, notable discrepancies persist between numerical and analytical studies for certain GNY models:
$\epsilon$-expansion analyses of the Gross–Neveu–XY and SO$(4)$-symmetric models—where Dirac fermions couple to two and six real bosonic fields, respectively—have suggested that the corresponding critical points may become unstable~\cite{GNXY_Hawashin,GNSO4_Uetrecht}, whereas quantum Monte Carlo simulations indicate continuous phase transitions \cite{GN_QMC4,GN_QMC5}.

Recently, however, a large-$N$ framework has been developed that provides a fully solvable generalization of the GNY model, enabling controlled access to the strong coupling Dirac quantum criticality \cite{DFS}.
In the conventional large-$N$ limit of the GNY model, the number of fermion flavors $N$ is taken to infinity while the number of bosons remains $\mathcal{O}(1)$, suppressing the renormalization from the bosonic fluctuations to the fermions.
In contrast, Ref.~\cite{DFS} introduced a large number $M$ of bosonic flavors and kept their ratio to fermions, $\gamma = M / (n_s N)$, fixed in the large-$N$ limit.
This construction captures the strong mutual renormalization between fermions and bosons at the resulting critical point, where the fermion scaling dimension is continuously controlled by the boson-to-fermion ratio $\gamma$.

\begin{figure}
    \centering
    \includegraphics[width = .9\columnwidth]{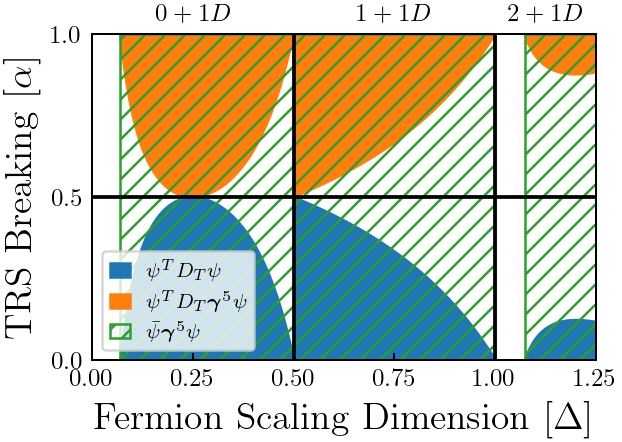}
    \caption{
    Zero temperature phase diagram for the generalized GNY model, as a function of $\alpha$, the degree of time-reversal symmetry $\trs$ breaking, and the fermion scaling dimension $\Delta$.
    Time reversal symmetry is preserved at $\alpha = 0, 1$, and maximally broken at $\alpha = 1/2$.
    The fermion scaling dimension depends on the dimensionality $D$ and the ratio of bosons to fermions $\gamma$.
    Colored regions indicate instabilities: blue and orange denote $s$-wave superconductivity where $\psi^T D_T \psi$ and $\psi^T D_T \gammaa_5 \psi$ respectively condense.
    Green denotes charge neutral instabilities, where $\bar\psi \gammaa_5 \psi$ condense.
    }
    \label{fig:alphac}
\end{figure}

In this work, we leverage the solvability of the generalized GNY model of Ref.\cite{DFS} together with conformal field theory to explore the various instabilities harbored by its critical point.
Our main results are summarized in Fig.~\ref{fig:alphac}, which presents the full zero-temperature phase diagram of superconductivity and charge neutral order that destabilize the critical point, as a function of the boson-to-fermion ratio $\gamma$ and the degree of time-reversal-symmetry (TRS) breaking $\alpha$.

In the inversion invariant representation relevant for physical lattice systems, we find a spontaneous breaking of chiral symmetry and a condensation of $\bar{\psi}\gammaa^5\psi$.
In $1+1$D this charge neutral instability is always present, whereas in $0+1$D and $2+1$D it emerges only under sufficiently strong fermion renormalization -- namely, for large boson-to-fermion ratio $\gamma$ or large fermion scaling dimension $\Delta$.

Conversely, the superconducting instabilities can be tuned by another parameter, $\alpha$, which controls the degree of time-reversal-symmetry (TRS) breaking.
In a manner reminiscent of Anderson's theorem \cite{Anderson}, superconductivity is most robust when TRS is preserved ($\alpha=0,1$) and vanishes when TRS is maximally broken ($\alpha=\frac12$).
Remarkably, we find that stronger fermion renormalization -- corresponding to larger $\gamma$ -- enhances superconductivity, as reflected in a larger critical value $\alpha_c$ up to which superconductivity remains stable.

\section{Model}
The model that we study is a generalization of \cite{DFS}, with inversion invariance and a parameter $\alpha$ that controls the degree of TRS breaking.
It consists of large number of flavors of massless Dirac fermions $\psi_{i}$, $i = 1, 2, \cdots N$ and bosons $\phi_n$, $n= 1, 2, \cdots M$ in spacetime dimensions $1\le D \le 3$ coupled through a translationally invariant but flavor-random Yukawa interaction, as given by the following Lagrangian,
\begin{equation}
    \mathcal{L} = \sum_{i=1}^N \bar\psi_i \slashed \partial \psi_i + \sum_{n=1}^{M = \gamma N}\frac{\phi_n^2}{2} + \sum_{ij,n} g_{ij}^n \bar\psi^i \psi^j \phi_n \,.
    \label{eq:Lagrangian}
\end{equation}
$\slashed{\partial} = \gammaa^\mu \partial_\mu, \ \bar\psi = \psi^\dagger \gammaa^0$, and $\gammaa^\mu\,, \ \mu=0,1,\cdots d$ are $n_s\times n_s$ gamma matrices that satisfy the Euclidean Clifford algebra, $\{\gammaa^\mu, \gammaa^\nu\} = 2\delta_{\mu\nu}$.
For $2+1$D, we employ a inversion-invariant four-component representation ($n_s = 4$) that admits a $\gammaa^5$ matrix.
This renders the formulation directly relevant to physical Dirac systems such as graphene, where parity symmetry relates the two Dirac cones.
Although a boson kinetic term of the form $b(\partial \phi_n)^2$ could in principle be included, it is irrelevant in the infrared \cite{DFS} and will be omitted.

The Yukawa coupling coefficient $g_{ij}^n = {g'_{ij}}^n + \im {g''_{ij}}^n$ is a random complex number, composed of a real symmetric part ${g'_{ij}}^n = {g'_{ji}}^n$ and an imaginary anti-symmetric part ${g''_{ij}}^n = -{g''_{ji}}^n$ due to Hermiticity.
They are zero mean, and their variance are given by,
\begin{equation}
\begin{split}
    & \overline{{g'_{ij}}^n {g'_{kl}}^m} = (1-\alpha) \frac{\rm{g}^2}{N^2} \delta_{nm} (\delta_{ik} \delta_{jl} + \delta_{il} \delta_{jk}), \\
    & \overline{{g''_{ij}}^n {g''_{kl}}^m} = \alpha \frac{\rm{g}^2}{N^2} \delta_{nm} (\delta_{ik} \delta_{jl} - \delta_{il} \delta_{jk})\,,
    \label{eq:disorderdef}
\end{split}
\end{equation}
where $\rm{g}$ denotes the strength of the Yukawa interaction.
In $2+1$D, a relativistic Mott transition \cite{Nambu} occurs at a critical coupling $\rm{g}_c$, where the fermions spontaneously acquire a Dirac mass.
By contrast, in $0+1$ and $1+1$D the model flows to criticality at low temperatures without fine tuning \cite{DFS}.
In what follows we concentrate on this critical regime, which not only provides access to the most interesting physics—the emergence of pairing from incoherent quasiparticles—but also permits the use of the powerful tools of conformal field theory.

The parameter $\alpha\in[0,1]$ interpolates between purely real and purely imaginary couplings.
For $\alpha=0$ or $1$ the system preserves time-reversal symmetry $\mathcal{T}$, while for intermediate values $\mathcal{T}$ is broken \footnote{For $\alpha = 0, 1$, the presence of $\trs$ can be seen explicitly after integrating out the bosonic fields $\phi_n$. We also note that even away from $\alpha = 0, 1$, the composite symmetry $\mathcal{T} \times \mathcal{PH} \times \mathbb{Z}_2$ remains, where $\mathcal{PH}$ denotes charge conjugation, and $\mathbb{Z}_2$, $\phi \rightarrow -\phi$.}.
The choice used in Ref.~\cite{DFS}, where $g_{ij}^n$ was drawn from a GUE ensemble, corresponds to $\alpha=1/2$, i.e. maximal breaking of $\mathcal{T}$.
As we will show, in a manner reminiscent of Anderson's theorem \cite{Anderson}, superconductivity is most robust when $\trs$ is preserved, and gradually weakens as $\trs$ is broken and vanishes completely at $\alpha = 1/2$.

Last, $\gamma = \frac{M}{n_s N}$ is an $\mathcal{O}(1)$ parameter quantifying the ratio of bosonic to fermionic flavors.
At the low temperature quantum critical point, $\gamma$ controls the scaling of the theory, and the Euclidean Green’s functions for bosons and fermions take the form \cite{DFS},
\begin{equation}
\begin{split}
    & G(\x) = \frac1N\sum_{i=1}^N\left<\psi_i(\x)\bar\psi_i(0) \right> \simeq A \slashed{\x}|\x|^{-2\Delta-1} \,, \\
    & F(\x) = \frac1M\sum_{n=1}^{M} \left<\phi_n(\x)\phi_n(0) \right> \simeq A_F |\x|^{4\Delta-2D} \,, \\
    & \textrm{ where } A_F = -\frac{C_{2\Delta-D/2} }{(2\pi)^D n_s \rm{g}^2 A^2 C_{2\Delta}} \,.
    \label{eq:greens}
\end{split}
\end{equation}
where the fermion scaling dimension $\Delta$ is determined by $\gamma$ through,
\begin{equation}
\begin{split}
    & \gamma = -\frac{B_{D/2-\Delta}C_{2\Delta}}{B_{D-\Delta}C_{2\Delta-D/2}} \\
    & C_a = (4\pi)^{D/2} \frac{\Gamma(D/2-a)}{2^{2a} \Gamma(a)}\,, \ B_a = \frac{C_{a-1/2}}{1-2a} \,,
    \label{eq:gamma}
\end{split}
\end{equation}
where the fermion scaling dimension $\Delta$ satisfies,
$\frac{D-1}2< \Delta < \min \left\{\frac{D}2, \frac{D+2}{4} \right\}$ \cite{DFS}.
In $0+1$D, Eq.\eqref{eq:gamma} admits two distinct solutions for $\Delta$ \cite{Bi:2017yvx,esterlis,Kim:2019lwh}.
These correspond to two separate universality classes distinguished by whether the effective interaction is attractive or repulsive.
The model defined in Eq.~\eqref{eq:Lagrangian} belongs to the attractive class, as can be seen by integrating out the bosons, which yields a negative-definite interaction, $H_{int} = - \frac12 \sum_n Q_n^2$, with $Q_n = \sum_{ij} g_{ij}^n \bar \psi_i \psi_j$.
This leads to a low temperature critical point with $\Delta > 1/4$.
In contrast, taking the interaction to be repulsive, $H_{int} = \frac12 \sum_n Q_n^2$, produces a critical point with $\Delta < 1/4$ \footnote{This can be implemented by taking the coupling to be anti-Hermitian, through $g_{ij}^n \rightarrow \im g_{ij}^n$. Note that integrating out the bosons restores Hermiticity in the effective fermionic theory.}.
We note that this distinction arises only in $0+1$D, where the model flows to criticality at low temperatures regardless of the sign of the interaction.

In what follows, we analyze the instabilities toward charge neutral order and superconductivity that destabilize the conformal saddle point discussed above, using the exact solvability of the model and conformal field theory.

\section{Diagnosis of Instability}
To characterize the possible instabilities, we first classify the fermion bilinears that can condense while preserving Lorentz invariance.
By Lorentz symmetry, the theory can only admit scalar and pseudoscalar condensates.

The order parameters to superconductivity are given by the charged fermion bilinears
$\cS^{\pm} = \psi^T \cs^\pm \psi$, where $\cs^\pm$ are antisymmetric matrices $D_T$ and $D_T \gammaa^5$, respectively.
Here, $D_T$ is the anti-symmetric charge conjugation matrix defined by $D_T \gammaa^\mu D_T^{-1} = -(\gammaa^\mu)^T$.
On the other hand, charge-neutral condensates are described by the bilinears
$\cM^\pm = \bar{\psi} \cm^\pm \psi$, where $\cm^+ = \Id$ and $\cm^- = \gammaa^5$
\footnote{In $2+1D$, a reducible representation with $\gammaa^5$ also admits a $\gammaa^3$; both anticommute with all $\gammaa^\mu$ and hence play equivalent roles in distinguishing charge neutral scalars and pseudoscalars. We therefore treat them as equivalent until the final discussion.}.

\begin{figure}[!htbp]
    \centering
    \includegraphics[width=1\linewidth]{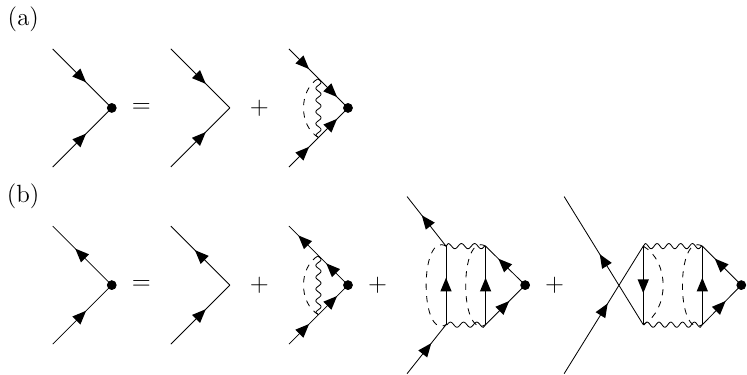}
    \caption{Bethe-Salpeter equations for the three point functions with (a)
    the charged fermion bilinear $\cS = \psi^T \cs \psi$, and (b) the charge neutral fermion bilinear $\cM = \bar\psi \cm \psi$. The straight lines denote fermion propagators, squiggly lines, boson propagators, and the dashed, disorder averaging. A dot indicates a fully dressed three-point function.}
    \label{fig:BS}
\end{figure}

To identify which of these channels becomes unstable, we proceed by contradiction: we assume that the ground state remains at the conformal saddle point of Eq.~\eqref{eq:greens} and compute the scaling dimensions of the fermion bilinears listed above.
As we will show, certain fermion bilinears develop complex scaling dimensions.
Although a complex scaling dimension strictly indicates only the inconsistency of the conformal ansatz, it has been conjectured that it corresponds to the condensation and finite expectation value of the associated fermion bilinear \cite{Kim:2019upg}, which in this case manifests as either superconductivity or charge neutral order.

To determine the scaling dimensions $h_{\cS,\cM}$ of the fermion bilinears $\cS^\pm$, $\cM^\pm$, we introduce the three-point functions $u$ and $v$ between the fermion bilinears and two dirac fermions,
\begin{align*}
& u^s_{\sigma \sigma'}(\x_1, \x_2; \x_0) = \braket{\bar\psi_{\sigma}(\x_1) \bar\psi_{\sigma'}(\x_2) \cS(\x_0)} \\
& v^m_{\sigma \sigma'}(\x_1,\x_2; \x_0) = \braket{\bar\psi_{\sigma}(\x_1) \psi_{\sigma'}(\x_2) \cM(\x_0)}
\end{align*}
Conformal invariance constrains the three point functions to scale as $|\x_{01}|^{-h_{a}} |\x_{02}|^{-h_{a}} |\x_{12}|^{-2\Delta+h_{a}}$, with $a = \cS, \cM$.
Taking the bilinear insertion point $\x_0$ far away from the origin, the correlators asymptotically reduce to,
\begin{subequations}
\label{eq:uv}
\begin{align}
    u^s_{\sigma_1 \sigma_2}(\x_{12}) & \simeq \frac{ c_s \cs_{\sigma_1 \sigma_2}}{|\x_{12}|^{2\Delta-h_\cS}} \,,
    \label{eq:v_S} \\
    v^m_{\sigma_1 \sigma_2}(\x_{12}) & \simeq \frac{c_m \cm_{\sigma_1 \sigma_2}}{|\x_{12}|^{2\Delta - h_\cM}}
    \label{eq:um}
\end{align}
\end{subequations}
where $c_{s,m} \propto |\x_0|^{-2h_a}$ are proportionality constants.

Crucially, the three-point functions can be self-consistently determined by the Bethe-Salpeter equations, shown in Fig.~\ref{fig:BS}.
In the IR, the bare diagrams on the right hand side are irrelevant, and they reduce to,
\begin{widetext}
\begin{subequations}
\label{eq:BS_all}
\begin{align}
   &  u_{\sigma_1 \sigma_2}(\x_1, \x_2) = \sum_{\sigma_3 \sigma_4} \int_{\x_3, \x_4} K^{sc}_{\sigma_1 \sigma_2; \sigma_3 \sigma_4}(\x_1,\x_2; \x_3, \x_4) u_{\sigma_3 \sigma_4}(\x_3, \x_4) \,,
   \label{eq:BS_sc} \\
    &\ \ \textrm{ where } K^{sc}_{\sigma_1 \sigma_2; \sigma_3 \sigma_4}(\x_1,\x_2; \x_3, \x_4) = -\gamma (1-2\alpha) \rm{g}^2 G_{\sigma_3 \sigma_1}(\x_{31}) G_{\sigma_4 \sigma_2}(\x_{42}) F(\x_{34})
    \notag \\
    & v_{\sigma_1 \sigma_2}(\x_1, \x_2) = \sum_{\sigma_3 \sigma_4} \int_{\x_3, \x_4}  K^n_{\sigma_1 \sigma_2; \sigma_3 \sigma_4}(\x_1,\x_2; \x_3, \x_4) v_{\sigma_3 \sigma_4}(\x_3, \x_4) \,, \textrm{ where } K^n = K^f + K^b \,,
    \label{eq:BS_n} \\
    &\ \ K^b_{\sigma_1\sigma_2; \sigma_3 \sigma_4}(\x_1,\x_2; \x_3, \x_4) = \gamma \rm{g}^2 G_{\sigma_2 \sigma_3}(\x_{23}) G_{\sigma_4 \sigma_1}(\x_{41}) F(\x_{34}) \,, \notag \\
    &\ \ K^f_{\sigma_1\sigma_2; \sigma_3 \sigma_4}(\x_1, \x_2; \x_3, \x_4) = \gamma \rm{g}^4 G_{\sigma_3\sigma_4}(\x_{34}) \int_{\x_{5},\x_{6}}  F(\x_{35}) F(\x_{46}) G_{\sigma_2 \sigma_5}(\x_{25}) G_{\sigma_5 \sigma_6}(\x_{56}) G_{\sigma_6 \sigma_1}(\x_{61}) + \x_5, \sigma_5 \leftrightarrow \x_6, \sigma_6 \,. \notag
\end{align}
\end{subequations}    
\end{widetext}
Here the kernels $K^{sc}$ and $K^n$ denote the insertion of a ladder rung to the three-point functions $u$ and $v$ respectively.
While the superconducting kernel $K^{sc}$ consists of a single boson ladder rung, that of the normal kernel $K^n$ consists of a sum of a boson ladder rung $K^b$ and a fermion ladder rung $K^f$.

Crucially, Eq.\eqref{eq:BS_all} indicates that the three point functions $u$ and $v$ are eigenfunctions to the kernels $K^{sc}$ and $K^n$, respectively, with unit eigenvalues.
Indeed, inserting Eqs.\eqref{eq:greens} and \eqref{eq:uv} to Eq.\eqref{eq:BS_all}, we find that $u$ and $v$ are eigenfunctions, with the eigenvalues,
\begin{subequations}
\label{eq:gs}
\begin{align}
    & g_{\cS^\pm}(h_{\cS^\pm}) =  \pm (1-2\alpha) k_s(h_{\cS^\pm}) \,, \label{eq:gs}  \\
    & g_{\cM^\pm}(h_{\cM^\pm}) =  \mp k_s(h_{\cM^\pm}) \,, \label{eq:gn}  \\
    & \textrm{ where } k_s(h) = \frac{C_{D-\Delta-h/2} C_{D/2-\Delta+h/2}}{B_{D-\Delta} B_{D/2-\Delta}} \,. \notag
\end{align}
\end{subequations}
Therefore, the scaling dimensions of the bilinears are determined by solving for $g_{a}(h_{a}) = 1$ for $a = \cS^\pm, \cM^\pm$, respectively.

\begin{figure}
    \centering
    \includegraphics[width=1.0\columnwidth]{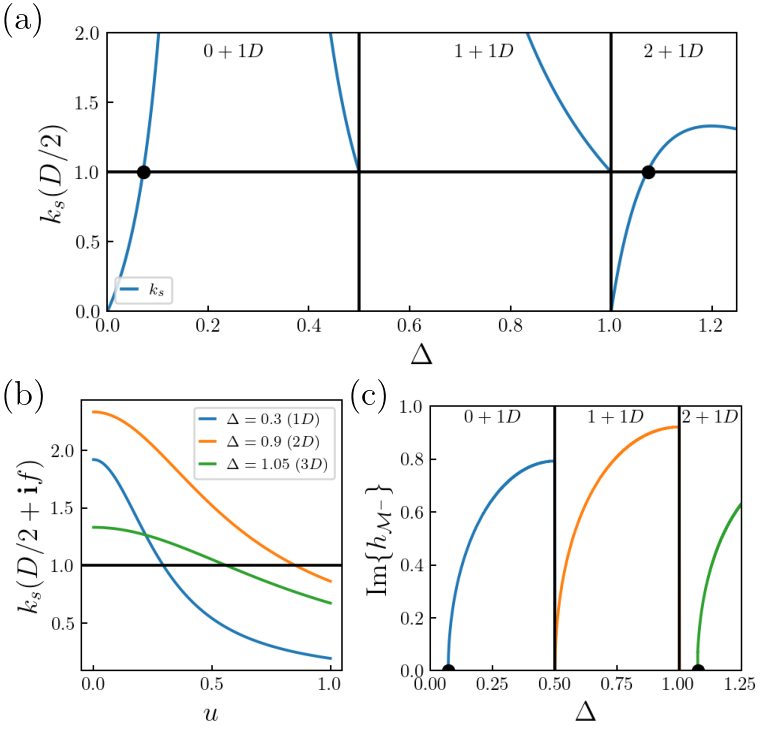}
    \caption{(a) $k_{s}(D/2)$ versus $\Delta$.
    $k_s(D/2) > 1$ for $0+1$D at $\Delta > \Delta_0 = 0.07088$, at all $\Delta$ in $1+1$D, and for $2+1$D at $\Delta > \Delta_2 = 1.07314$. $\Delta_{0,2}$ are marked by black dots.
    (b) The operators $\cM^-$ (or $\cS^\pm$ for $\alpha = 0,1$) acquire complex scaling dimensions $D/2+\im f$, where $k_s(D/2 + i f) = 1$. (c) The imaginary component of the scaling dimension $\rm{Im}\{h_{\cM^-}\}$ is shown.}
    \label{fig:k_sp}
\end{figure}

Since our goal is to assess the stability of the conformal critical point, let us assess whether $h_{a}$ is complex.
Two properties of $k_s(h)$ simplify this analysis.
First, $k_s(h)$ is real if $h$ is real or of the form $h = D/2 + \im f$, and we therefore focus on the latter.
Second, $k_s(D/2+\im f)$ is positive definite and monotonically decreasing as a function of $|f|$, decaying to zero as $|f| \rightarrow \infty$ (See Fig.\ref{fig:k_sp}b).

We first consider charge neutral instabilities.
From the two properties of $k_s$ discussed above, it is clear that $g_{\cM^+}(D/2+\im f) < 0$, excluding complex solutions to $h_{\cM^+}$.
In contrast, when $k_s(D/2) > 1$, the operator $\cM^-$ obtains a complex scaling dimension.
As shown in Fig.\ref{fig:k_sp}a) this condition is satisfied in $1+1$D, and also in $0+1$D and $2+1$D once the fermion scaling dimension exceeds $\Delta_{0} = 0.07088$ and $\Delta_{2} = 1.07314$, respectively.
In this regime, $h_{\cM^-}$ develops a nonzero imaginary part, as illustrated in Fig.\ref{fig:k_sp}b) and c).
This signals an instability of the conformal critical point, suggesting that the system evolves into a gapped phase characterized by condensation of $\cM^-$.

Interestingly, this instability arises only in the regime of strong fermion renormalization, above a critical boson-to-fermion ratio $\gamma_c$.
At $\gamma = 0$, the fermions are free and their scaling dimension takes the bare value $(D - 1)/2$ for the (repulsive) $0+1$D and $2+1$D
\footnote{In the attractive cases of both $0+1$D and $1+1$D, as $\gamma \rightarrow 0$ the bare fermion scaling dimension approaches $D/2$, since the softest bosons condense and the fermions spontaneously acquire a mass term.}.
As the number of bosons increase, the fermions become increasingly renormalized, acquiring a larger anomalous dimension.
The onset of the instability, defined by $k_s(D/2) > 1$, occurs at $\gamma > \gamma_{c0} = 0.1295$ in the repulsive $0+1$D and $\gamma > \gamma_{c2} = 0.3729$ in $2+1$D, as determined from Eqs.\eqref{eq:gamma}.
This behavior is reminiscent of $\epsilon$-expansion results of the Gross–Neveu–XY and SO(4) symmetric models, which found the critical point to be stable only above a critical number of fermion flavors \cite{GNXY_Hawashin,GNSO4_Uetrecht}.


The physical interpretation of the condensation of $\cM^-$ depends on how the components of the Dirac spinor are embedded into microscopic lattice degrees of freedom, such as sublattice, valley, or spin.
In $1+1$ dimensions, a lattice formulation in terms of slowly varying chiral fermions at half filling identifies the bilinear $\cM^-$ with a charge-density-wave order parameter.
In $2+1$ dimensions, Dirac fermions on the hexagonal lattice provide a natural realization of the generalized Gross–Neveu–Yukawa model;
however, the physical meaning of the condensation of $\cM^-$ is dictated by the microscopic details of how the interaction $(\bar{\psi}\psi)^2$ is embedded into the lattice Hilbert space and can correspond to sublattice symmetry breaking or Kekulé bond ordered phases.

We now turn to superconducting instabilities.
Using the same analysis as above, we find that when $|1 - 2\alpha| k_s(D/2) > 1$, one of the two fermion bilinears $\cS^\pm$ acquires a complex scaling dimension:
$\cS^+$ for $\alpha < 1/2$, and $\cS^-$ for $\alpha > 1/2$.
This signals condensation of $\cS$ and the emergence of an $s$-wave superconducting ground state.
Accordingly, the critical degree of time-reversal-symmetry (TRS) breaking at which $s$-wave superconductivity sets in is given by
\begin{equation}
\alpha_c = \frac{1}{2} \pm \frac{1}{2 k_s(D/2)} \,.
\label{eq:alphac}
\end{equation}

Eq.\eqref{eq:alphac} yields the superconducting phase boundaries of Fig.\ref{fig:alphac}, with blue indicating the condensation of $\cS^+$, and orange, that of $\cS^-$.
Superconductivity is strongest when $\trs$ is preserved at $\alpha = 0$ or $1$, albeit only above the critical fermion scaling dimensions $\Delta_{0,2}$ in $0+1$D and $2+1$D, respectively.
Interestingly, stronger fermion renormalization -- corresponding to a larger fermion to boson ratio $\gamma$ -- enhances pairing, as reflected in the increased critical value of TRS breaking $\alpha_c$
\footnote{Note that in $0+1$D, the limit $\gamma \rightarrow \infty$ corresponds to $\Delta = 1/4$.}.

We find excellent agreement between our results and previous studies of related models through the Eliashberg equations.
Ref.\cite{sss1} recently studied a closely related model in $2+1$D with real coupling coefficients ($\alpha = 0$), and observed that superconductivity appears only for fermion scaling dimensions $\Delta > \Delta_2$.
In addition, Ref.\cite{2020Hauck} studied a comparable model in $0+1$D with attractive interactions and a boson-to-fermion ratio $\gamma = \frac12$ (corresponding to $\Delta = 0.4203$) and obtained $\alpha_c = 0.312 \pm 0.002$ from numerical solutions of the Eliashberg equations.
Our analytical expression, Eq.\eqref{eq:alphac} gives $\alpha_c = 0.313453$, in excellent agreement.

\section{Discussion}
In this paper, we categorically analyzed the instabilities to a generalized Gross–Neveu-Yukawa critical point.
Starting from the assumption that the ground state is conformal, we employed conformal field theory and the large-$N$ Bethe–Salpeter equations to compute the scaling dimensions of various fermion bilinears.
We identified several bilinears whose scaling dimensions become complex, providing—by proof of contradiction—evidence that the conformal phase is unstable.
We conjecture that, at zero temperature, the gapped ground state is characterized by spontaneous symmetry breaking and condensation of the fermion bilinear with the complex scaling dimension.
A complete characterization of the ordered phase—particularly in the presence of multiple competing instabilities—is left for future work.

Within our model, superconductivity can be tuned either by the boson-to-fermion ratio $\gamma$ or by the degree of time-reversal-symmetry (TRS) breaking $\alpha$.
It is most robust when TRS is preserved ($\alpha = 0, 1$) and vanishes when TRS is maximally broken ($\alpha = 1/2$).
Moreover, we find that superconductivity is generally enhanced at larger $\gamma$, where fermions experience stronger renormalization.
It would be interesting to see whether a holographic interpretation of this superconductivity, similar to Ref.\cite{Inkof_2022} could give an intuition to this result.

Interestingly, in $0+1$D we find superconductivity even when the interaction is positive definite and thus purely repulsive.
This suggests an unconventional mechanism in which superconductivity emerges from a sea of incoherent electrons through repulsive interactions alone -- entirely beyond the traditional BCS paradigm.
It would be worthwhile to test this scenario through exact-diagonalization studies and to examine whether the mechanism persists in higher dimensions.

\textit{Note added:} As we were finalizing this work, independent works appeared \cite{sss1,sss2}, which examines superconductivity by the Eliashberg equations in a similar model albeit with real coupling coefficients.
The results are in agreement where they overlap.

\acknowledgements
JK is grateful to Ehud Altman, Matthew O'Brien, Junyi Cao, Xiangyu Cao, Shubhayu Chatterjee, Eduardo Fradkin, Igor Herbut, Benjamin Moy, Grgur Palle, and J\"org Schmalian for helpful discussions and comments.
\bibliography{ref}

\end{document}